\documentclass[letterpaper,twocolumn,showpacs,floatfix,groupaddress,longbibliography]{revtex4-1}
\usepackage[latin1]{inputenc}
\usepackage{bm}
\usepackage{multirow}
\usepackage{amssymb}
\usepackage{amsbsy}
\usepackage{amsmath,  amsthm, amsfonts, mathrsfs}
\usepackage{graphicx}
\usepackage{epsfig}
\usepackage{placeins}
\usepackage{float}
\usepackage{trfsigns}
\usepackage[usenames,dvipsnames]{color}
\usepackage[colorlinks,linkcolor=blue,citecolor=blue,urlcolor=blue,breaklinks=true]{hyperref}

\makeatletter

\begin{document} 

 \title{Modeling the Impact of Hamiltonian Perturbations on Expectation Value Dynamics}
 
 \author{Robin Heveling}
 \email{rheveling@uos.de}
 \affiliation{Department of Physics, University of Osnabr\"uck, D-49069 Osnabr\"uck, Germany}

 \author{Lars Knipschild}
 \email{lknipschild@uos.de}
 \affiliation{Department of Physics, University of Osnabr\"uck, D-49069 Osnabr\"uck, Germany}

 \author{Jochen Gemmer}
 \email{jgemmer@uos.de}
 \affiliation{Department of Physics, University of Osnabr\"uck, D-49069 Osnabr\"uck, Germany}

\begin{abstract}
Evidently,  some relaxation dynamics, e.g. exponential decays, are much more common in nature than others. Recently there have been attempts to explain this observation on the basis of  ``typicality of perturbations'' with respect to their impact on expectation value dynamics. 
These theories suggest that a majority of the very numerous, possible Hamiltonian perturbations entail more or less the same type of alteration of the decay dynamics. Thus, in this paper, we study how the approach towards equilibrium in closed quantum systems is altered due to weak perturbations. To this end, we perform numerical experiments on a particular, exemplary spin system. We compare our numerical data to predictions from three particular theories. We find satisfying agreement in the weak perturbation regime for one of these approaches.
\end{abstract}


\maketitle

\section{Introduction}
\label{intro}
The issue of the apparent emergence of irreversible dynamics from the underlying theory of quantum mechanics still lacks an entirely  satisfying answer \cite{gogolin}.
While concepts like the ``eigenstate thermalization hypothesis'' \cite{deutsch,srednicki} or ``typicality'' \cite{lloyd,goldstein,reimann} hint at fundamental 
mechanisms ensuring eventual equilibration, they are not
concerned in which manner this equilibrium is reached. It is an empirical fact that some relaxation dynamics, e.g. exponential decays, occur much more often 
in nature that others, e.g. recurrence dynamics. There are efforts to attribute this dominance to a certain sturdiness of some dynamics against a large class of 
small alterations 
of the Hamiltonian \cite{stability}. In general, it is of course impossible to predict how the unperturbed dynamics will change due to an arbitrary  perturbation. 
However, theories aiming at capturing the typical impact of generic perturbations have recently been suggested.
\newline
\noindent
In the following three such theories that predict the altered dynamics due to weak, generic pertubations are very briefly presented. Notably, 
Refs. \cite{reimann1,richter1,stability} are 
concerned with describing the modified dynamics under certain assumptions, cf. also Sect. \ref{modpedy}. 
\newline
In \mbox{Ref. \cite{reimann1}} the authors consider an entire ensemble of ``realistic'' Hamiltonian pertubations, i.e. the ensemble members are sparse and possibly 
banded in 
the eigenbasis of the unperturbed Hamiltonian. The authors analytically calculate the ensemble average of time-dependent 
expectation values and argue firstly that 
the ensemble variance is small and secondly  that thus a perturbation of actual interest is likely a ``typical'' member of the ensemble. Calculating the ensemble average,
the authors eventually arrive at the result 
that the unperturbed dynamics will likely be exponentially damped with a damping factor scaling quadratically with the perturbation strength. 
\newline

\noindent
A similar random matrix approach is taken in Ref. \cite{richter1}. In this paper, the authors base their argument on projection operator techniques.  
Again, the ensemble of perturbations that are essentially random matrices  in the eigenbasis of the unperturbed Hamiltonian leads to an exponential damping of the unperturbed dynamics at sufficiently long times, with a damping constant scaling quadratically with the pertubation strength. Routinely, the specific projection operator technique (``time convolutionless'' \cite{breuer}) yields a time-dependent damping factor, which ensures that the slope of the time-dependent expectation value at $t=0$ remains unchanged by the perturbation.
\newline
Lastly, the authors of Ref. \cite{stability}, other than the authors of \mbox{Ref. \cite{reimann1,richter1}}, focus on the matrix structure of the perturbation 
in the eigenbasis of the observable rather than in the eigenbasis of the unperturbed Hamiltonian.
In this paper, the modified dynamics is not necessarily obtained by a direct damping, but rather by an exponential damping of the memory-kernel. As the predictions of this scheme are somewhat involved, we specifically outline them below in Sect. \ref{memkern}.\newline
\noindent
This paper is structured as follows. Firstly, in Sect. \ref{memkern} we give a short introduction to the memory-kernel ansatz employed in Ref. \cite{stability}. 
The numerical setup is described in  Sect. \ref{setup}.
In Sect. \ref{res} the numerical results from the solution of the Schr\"odinger equation are presented and discussed. Sect. \ref{modpedy} scrutinizes the 
possible application of the above three theories (Refs. \cite{reimann1,richter1,stability}) to the obtained numerical results and 
the accuracy of the respective predictions. Eventually, we conclude in Sect. \ref{sumcon}.
\section{Memory-Kernel ansatz}
\label{memkern}
To outline the memory-kernel ansatz we first need to introduce the general description of dynamics by means of integro-differential equations of the 
Nakajima-Zwanzig type \cite{kubo}.
Consider some (reasonably well-behaved) time-dependent function $a(t)$, e.g. the expectation value of an unitarily evolving observable. There exists a map between $a(t)$ and its so-called memory-kernel $K(\tau)$, implicitly defined by the integro-differential equation
\newpage
\begin{equation}
\label{mem}
\dfrac{\text{d}a(t)}{\text{d}t}= - \int_0^t K(t-t')a(t')\,\text{d}t' = -(K\ast a\,)(t)\,.
\end{equation}
This map is bijective, i.e., it is possible to calculate the memory-kernel $K(\tau)$ solely from the function $a(t)$ and, vice versa,
it is possible to calculate the function $a(t)$, given the memory-kernel $K(\tau)$ and some initial value $a(0)$. Broadly speaking, the memory-kernel 
describes how a system remembers its history. Ref. \cite{stability} now suggests that the generic impact of a certain class of perturbations is best captured  by 
describing its effect on the respective memory-kernel. If the perturbation $V$ is narrow-banded in the eigenbasis of the observable $A$, i.e. $[V,A]\approx 0$, then the 
memory-kernel $K(\tau)$ corresponding to the unperturbed dynamics will  be exponentially damped as
\begin{equation}
\label{damp}
\tilde{K}(\tau)=\text{exp}(-\gamma \tau)K(\tau)\,.
\end{equation} 
\noindent
To obtain the modified dynamics, we proceed as follows: from $a(t)$ we calculate the memory-kernel $K(\tau)$ 
and damp it according to Eq. \eqref{damp}. Plugging $\tilde{K}(\tau)$ back into Eq. \eqref{mem} and solving for $\tilde{a}(t)$ yields the modified dynamics.
In this procedure $\gamma$ is a free fit parameter. 
\begin{equation}
a(t) \rightarrow K(\tau)\rightarrow \tilde{K}(\tau)\rightarrow \tilde{a}(t)
\end{equation}
Note that, in the context of the below (cf. Sect. \ref{setup}) defined spin ladders, this is an heuristic approach. 
However, for other scenarios, this memory-kernel ansatz is proven to hold \cite{magic}. These scenarios feature systems for which the 
eigenstate thermalization hypothesis (ETH) \cite{srednicki} applies to some observable $A$. 
The role of the perturbation is taken by an environment, which induces pure dephasing in the 
eigenbasis of $A$. The memory-kernel ansatz then applies to the expectation value of $A$. The rationale behind using the memory-kernel ansatz in the context of, e.g., 
isolated spin ladders, is that a generic perturbation $V$  with $[V,A]\approx 0$ may have  an effect comparable to the above dephasing. Moreover, the applicability 
of the  memory-kernel ansatz to closed systems has been numerically demonstrated for some concrete but rather abstract examples in Ref. \cite{stability}. It has also been 
found to yield surprisingly accurate results for systems similar to the ones discussed below \cite{richtertransport}. 
\section{Setup}
\label{setup}
\begin{figure}[H]
        \centering
              \includegraphics[width=0.4\textwidth]{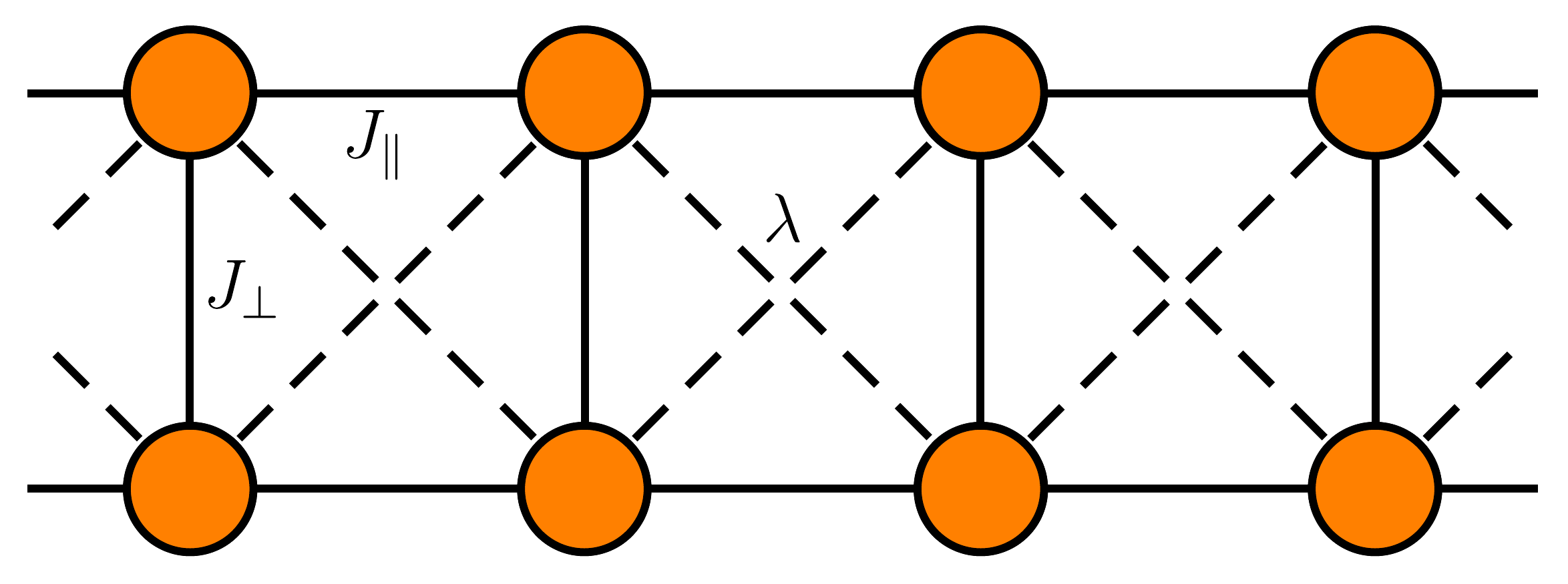}
            \caption{Orange circles mark spin sites, solid black lines mark Heisenberg interaction. Dashed diagonal lines indicate the pertubation.}
                \label{model1}
        \end{figure}

We consider a periodic spin-$1/2$ ladder described by the (unperturbed) Hamiltonian
    \begin{equation}
    H_0 = H_{\parallel} + H_{\perp}\,,
    \end{equation}
    with the chain Hamiltonian
 \begin{equation}
    H_{\parallel}=J_{\parallel} \sum_{k=1}^2 \sum_{l=1}^L \vec{S}_{l,k}\cdot\vec{S}_{l+1,k}
    \end{equation}
    and the rung Hamiltonian
    \begin{equation}
    H_{\perp}=J_{\perp} \sum_{l=1}^L \vec{S}_{l,1}\cdot\vec{S}_{l,2}\,,
    \end{equation}
where $\vec{S}_{l,k} = (S_{l,k}^x,S_{l,k}^y,S_{l,k}^z)$ are spin-$1/2$ operators on lattice site $(l,k)$ and $L+1 \equiv 1$. The interaction strength along the legs (rungs) is denoted by $J_{\parallel}$ ($ J_{\perp})$ and set to unity. Additional diagonal bonds act as a pertubation $V$, the parameter $\lambda$ indicates the pertubation strength. This results in the total Hamiltonian
\begin{equation*}
H = H_0 + \lambda V\,,
\end{equation*}
which is displayed in Fig. \ref{model1}. The observables of interest are the magnetizations on each rung, which are given by
\begin{equation}
S^z_l = S^z_{l,1}+S^z_{l,2}\,,
\end{equation} 
and the respective Fourier modes
 \begin{equation}
 \label{mod}
 S^z_q = \sum_{l=1}^L \cos[q(l-L/2)]S^z_l\,,
    \end{equation}
with discrete momenta $q=2\pi k/L$ with $k=0,1,...\,,L-1$.
We numerically solve the Schr\"odinger equation and study the dynamics of the time-dependent expectation values $p_l(t) =\langle S^z_l(t)  \rangle $ of the magnetization profile along the ladder
 as well as time-dependent expectation values $p_q(t) = \langle S^z_q(t)  \rangle $ of the Fourier modes, especially the slowest mode with $q=2 \pi/L$. In order to be able to clearly discriminate between predictions from the memory-kernel ansatz and the other two theories, we choose a perturbation with a specific, yet physically common  property named below. This pertubation on the diagonals of the ladder only consists of $S^zS^z$-terms.
        \begin{equation}
    V= \sum_{l=1}^L \big{(}S_{l,1}^z S_{l+1,2}^z+S_{l,2}^z S_{l+1,1}^z\big{)}
    \end{equation}
    In this manner, the observables of interest do commute with the pertubation, i.e. $[V,S^z_q] = 0$. In other words, the pertubation $V$ is diagonal in the eigenbasis
    of the observable. 

        \newpage
        \noindent
We consider two types of initial states. The first initial state is given by
    \begin{equation}
    \label{ini}
    \rho_1(0) \propto 1 - \varepsilon\,S^z_{L/2}\,,
    \end{equation}
    where $\varepsilon$ is a small, positive, real number.
This state can be regarded as the high temperature, strong magnetic field limit ($\beta \rightarrow 0$ while $ \beta B = \varepsilon$) of the Gibbs state    
    \begin{equation}
    \label{ini2}
    \rho_2(0) \propto \text{exp}[-\beta(H + B S^z_{L/2})]\,,
    \end{equation}
    which is the second initial state of interest. 
\section{Numerical Results on the perturbed dynamics}
\label{res}
\begin{figure}[b!]
    \centering
          \includegraphics[scale=0.28]{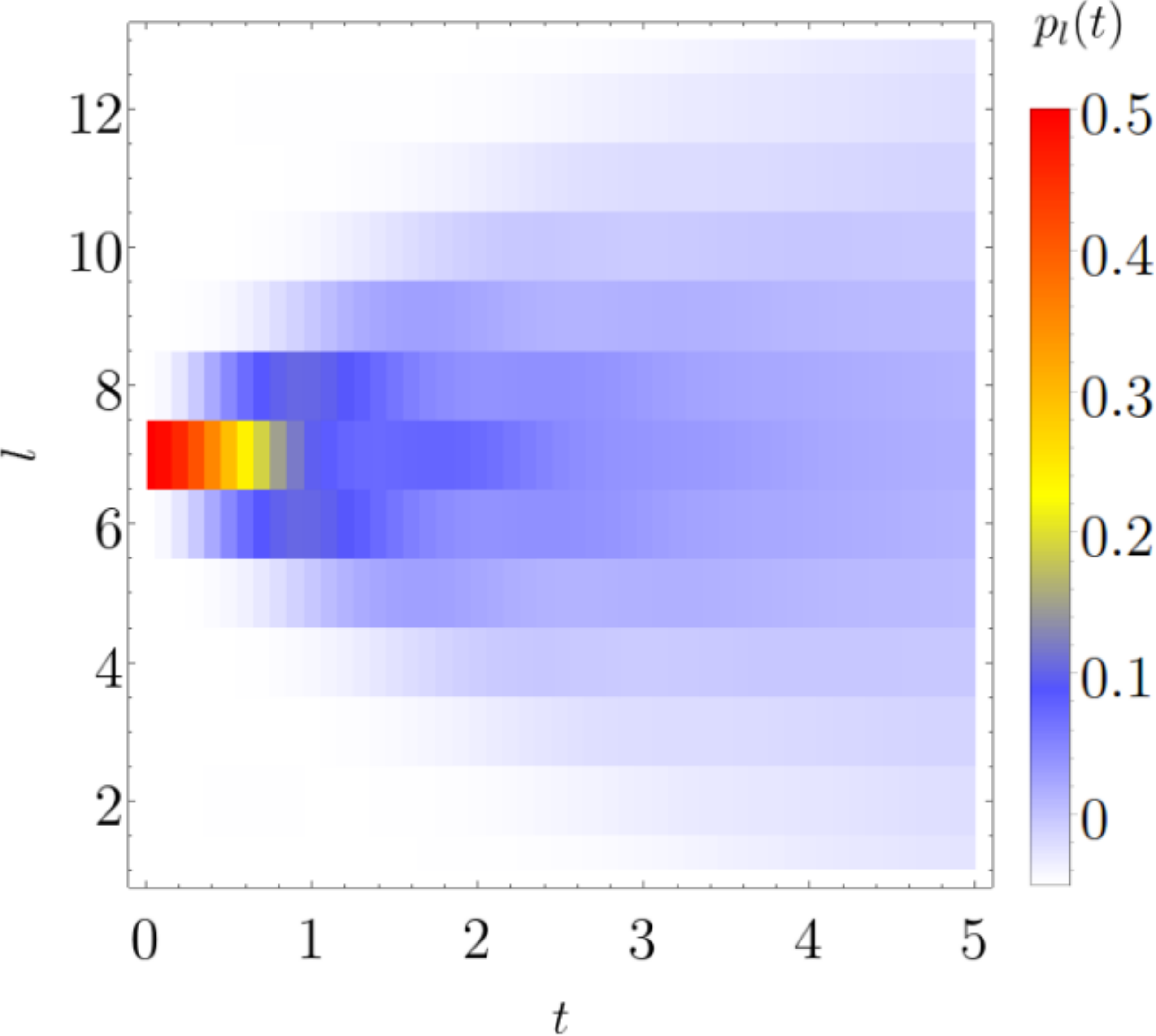}
        \caption{Broadening of the magnetization profile of a spin ladder with $L=13$ rungs prepared in the initial state $\rho_1(0)$ .}
    \label{wave}
    \end{figure}

We now present our numerical results. We prepare a spin ladder with $L=13$ rungs (i.e. $N=26$ spins) in the initial states mentioned aboved, which both feature a sharp magnetization peak in the middle of the ladder. During the real time evolution the magnetization will spread throughout the ladder diffusively \cite{richtertransport}, which can be seen in Fig. \ref{wave}.\newline

\begin{figure}[t!]
\centering
  	\includegraphics[width=0.43\textwidth]{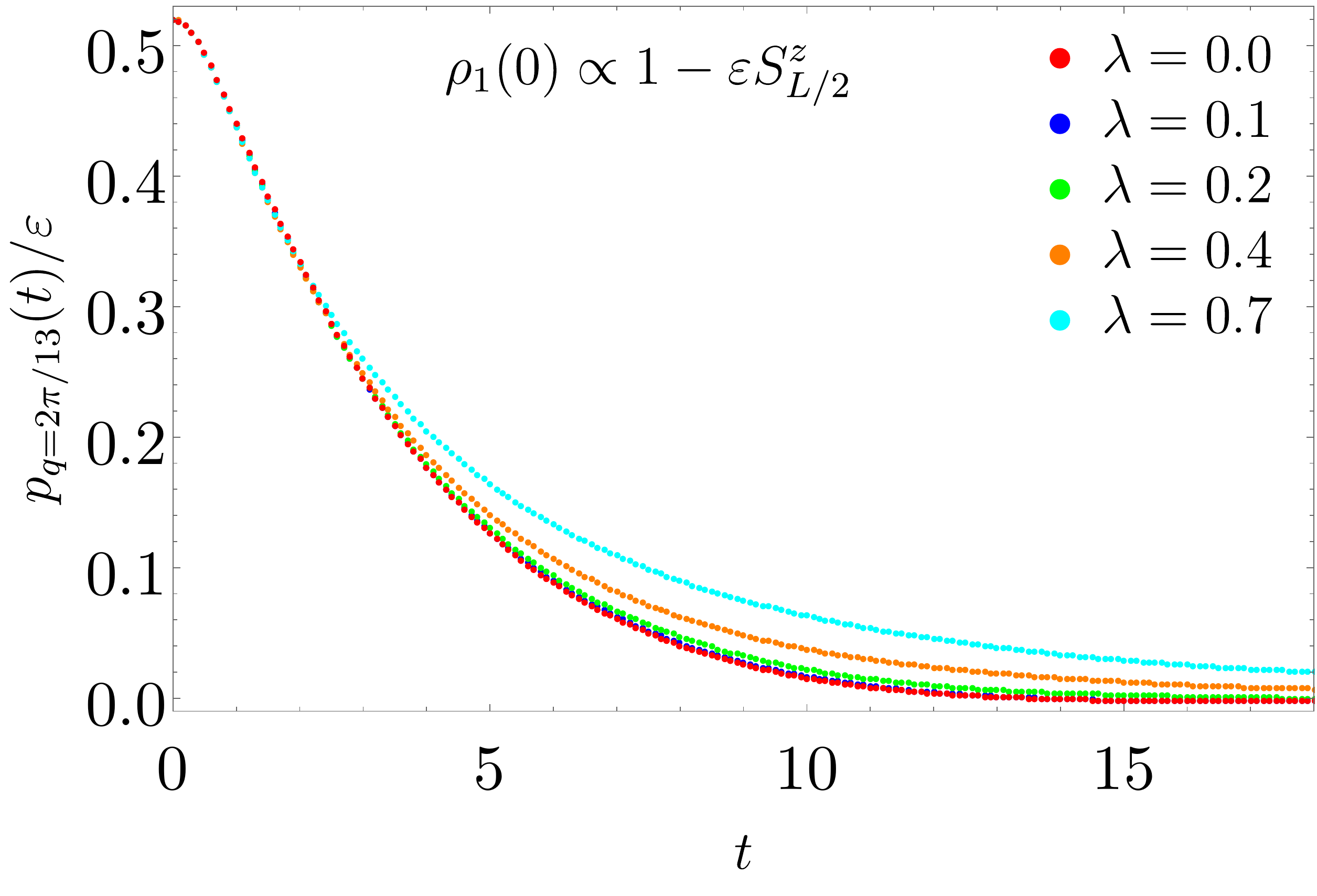}
	\caption{The time-dependence of the slowest mode with \mbox{$q = 2\pi/13$} is depicted for various pertubation strengths for the initial state $\rho_1(0)$. For small pertubations the unpertubed dynamic ($\lambda=0.0$, red curve) remains basically unchanged. For stronger pertubations there is a noticeable deviation.}
\label{dyn}
\end{figure}

\noindent
From Eq. \eqref{mod} we obtain the Fourier modes of the broadening process. We choose to investigate the slowest mode with $q=2 \pi/13$ in depth since it is closest to an exponential decay.  In Fig. \ref{dyn} the slowest mode is depicted for different pertubation strengths for the first initial state $\rho_1(0)$. The unperturbed dynamic (red curve, $\lambda=0.0$) remains basically unaltered by weak pertubations. Cranking up the pertubation strength (to $\lambda = 0.4$ or $\lambda =0.7$) leads to a noticable deviation and the equilibration process is much slower than in the weakly perturbed case.\newline
The same qualitative behavior remains when going to finite temperature $\beta = 0.1$ and finite magnetic field $B=5.0$.

\begin{figure}[b!]
\centering
  	\includegraphics[width=0.43\textwidth]{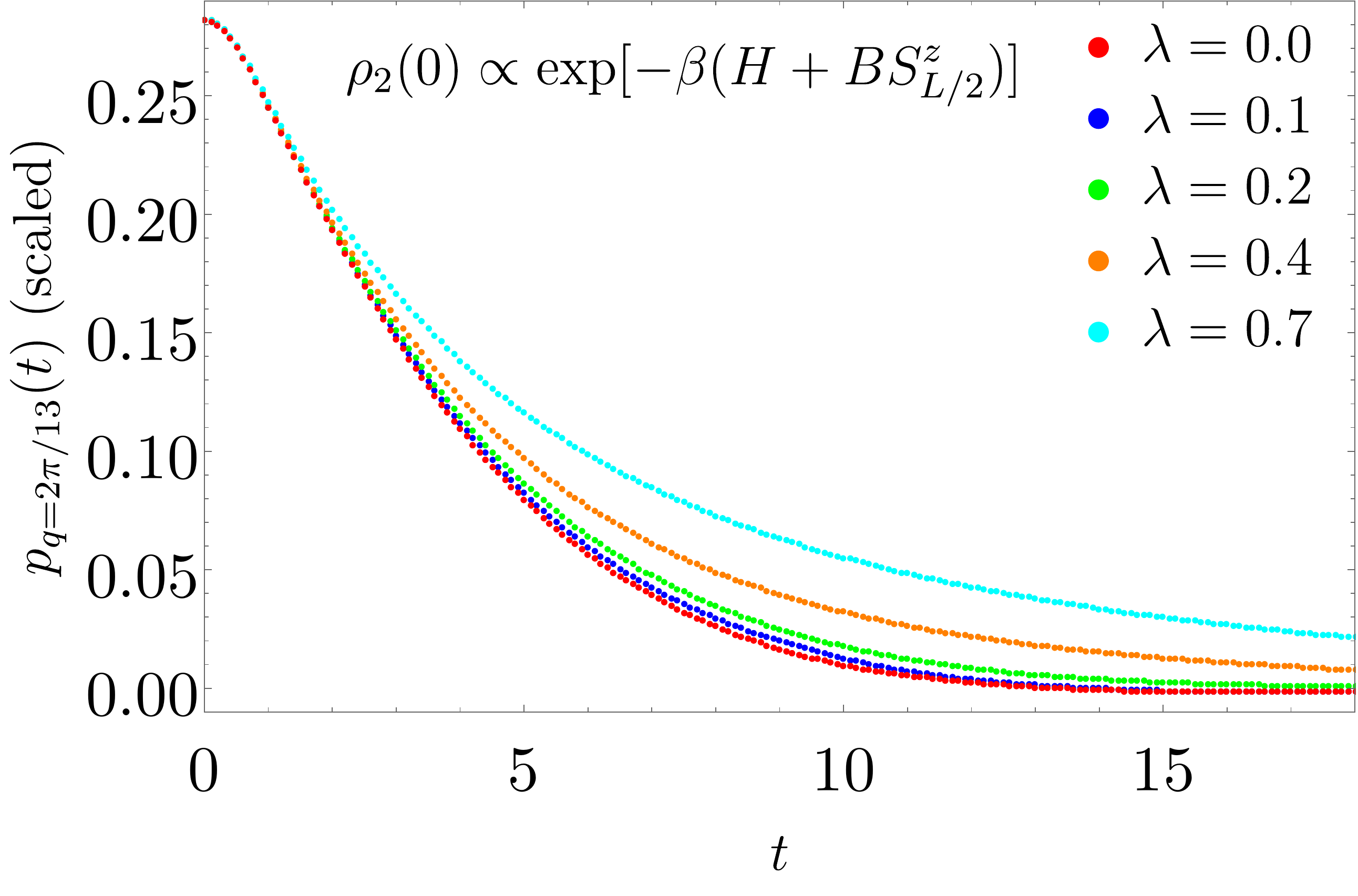}
	\caption{The time-dependence of the slowest mode with \mbox{$q = 2\pi/13$} is depicted for various perturbation strengths for a Gibbsian initial state $\rho_2(0)$ with $\beta=0.1$ and $B=5.0$. Similar behavior as for the first initial state can be observed.}
	\label{dyn2}
\end{figure}

\noindent
In this case, the initial value $p_q(0)$ depends on the pertubation strength $\lambda$ since the total Hamiltonian is part 
of the initial state $\rho_2(0)$, cf. Eq. \eqref{ini2}. To be able to compare the dynamics for various $\lambda$, the curves 
are scaled such that they all start at the same initial value of the unperturbed dynamic. The results are depicted in \mbox{Fig. \ref{dyn2}}.
For weak pertubations the deviation from the unperturbed dynamic is again small, although now clearly visible. For stronger pertubations the
dynamics equilibrate again more slowly, however, the discrepancy to the unperturbed dynamic is more severe compared to the first initial
state $\rho_1(0)$ in Fig. \ref{dyn}. A rough estimate indicates that at inverse temperature $\beta = 0.1$ the mean energy is down-shifted by approximately half a standard deviation of the full energy spectrum of the system with respect to the infinite temperature case ($\beta = 0$). Thus,  $\beta = 0.1$  is noticeable far away from infinite temperature while still not exhibiting low temperature phenomena.

\section{Modelling the perturbed dynamics}
\label{modpedy}
Is it possible to describe the observed behavior to some extend by any the three theories mentioned in the introduction? Before we present a somewhat bold, 
simple comparison of the predictions from said modeling schemes with the actual perturbed dynamics, we briefly comment on the agreement 
of our setup (cf. Sect. \ref{setup}) with the preconditions of the respective theories. \newline

\begin{figure}[t!]
\centering
  	\includegraphics[width=0.46\textwidth]{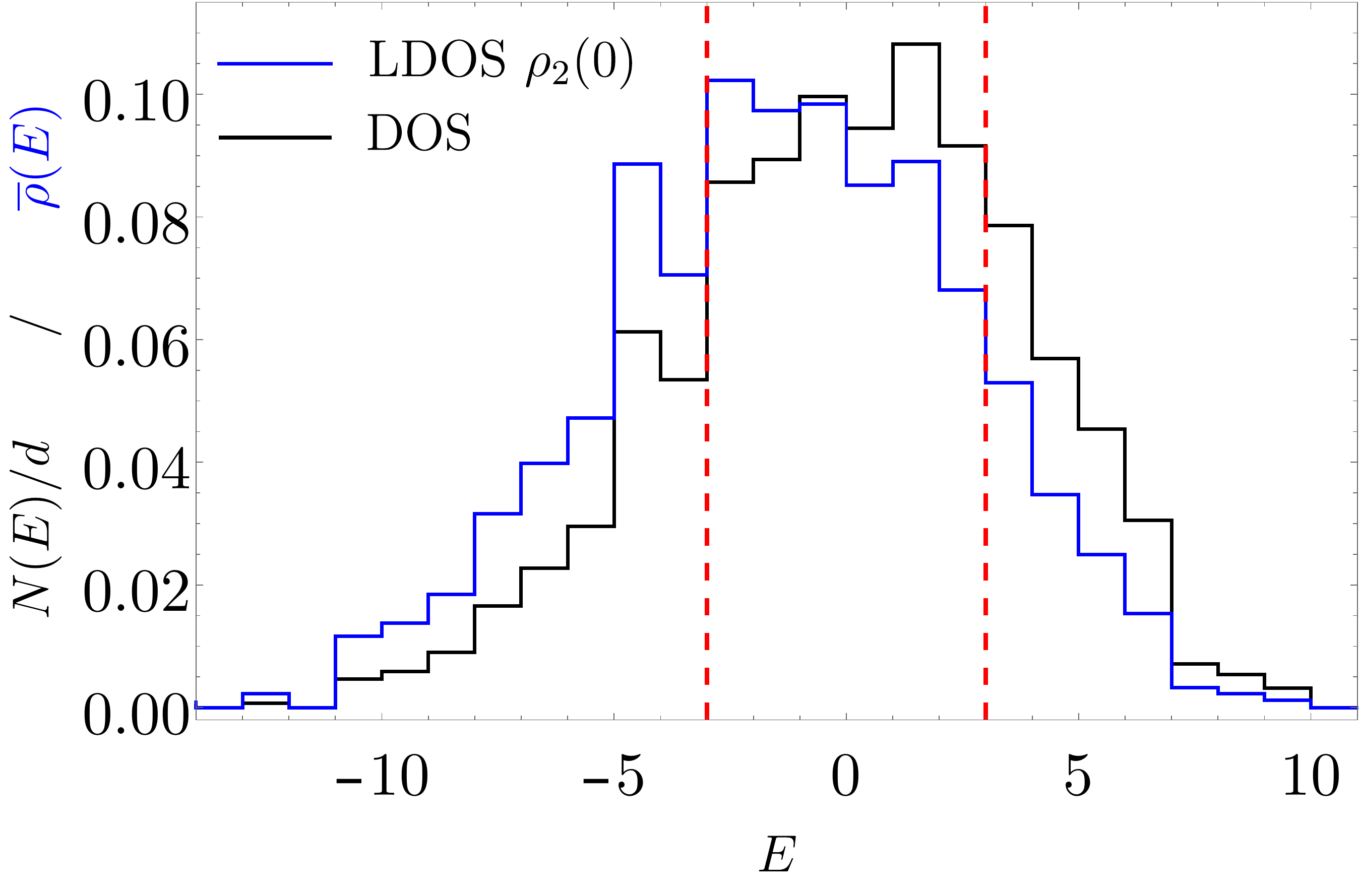}
	\caption{Density of states (black) for $N=12$ spins. The quantity $N(E)$ is the number of energy eigenstates in a particular bin of size one, $d=4096$ is the Hilbert space dimension. The interval of approximately constant DOS is marked from $E=-3\,\ldots\,3$ by red lines. A histogram of the local density of states (LDOS), i.e. the probability  to find the system at a certain energy, is shown in blue.  The quantity $\bar{\rho}(E)$ indicates the weight in a given bin. The LDOS of $\rho_1(0)$ is exactly identical to the DOS (black) and not shown again.}
	\label{dos}
\end{figure}

\noindent
The theory advocated in Ref. \cite{reimann1} relies on a constant density of states (DOS) within the energy interval occupied by the initial 
state $\rho(0)$ with respect to the unperturbed Hamiltonian $H_0$.First of all it should be noted that it is rather hard to 
check whether or not this criterion 
applies in standard situations with larger systems. However, a histogram corresponding to the DOS of $H_0$ for a ``small'' system with $N=12$ spins is
depicted in Fig. \ref{dos}. The red dashed vertical lines are intended to mark the regime of more or less constant DOS (of course this choice is rather arbitrary).
The initial state $\rho_1(0)$ populates the full spectrum with equal weight, i.e. $57 \%$ of the weight falls into the interval of approximately constant DOS.
Likewise, although populating more low-lying energy eigenstates, a large portion ($54 \%$) of the weight of the initial state $\rho_2(0)$ still falls into the 
interval of approximately constant DOS, cf. Fig. \ref{dos}. \\

\begin{figure}[b!]
\centering
  	\includegraphics[width=0.43\textwidth]{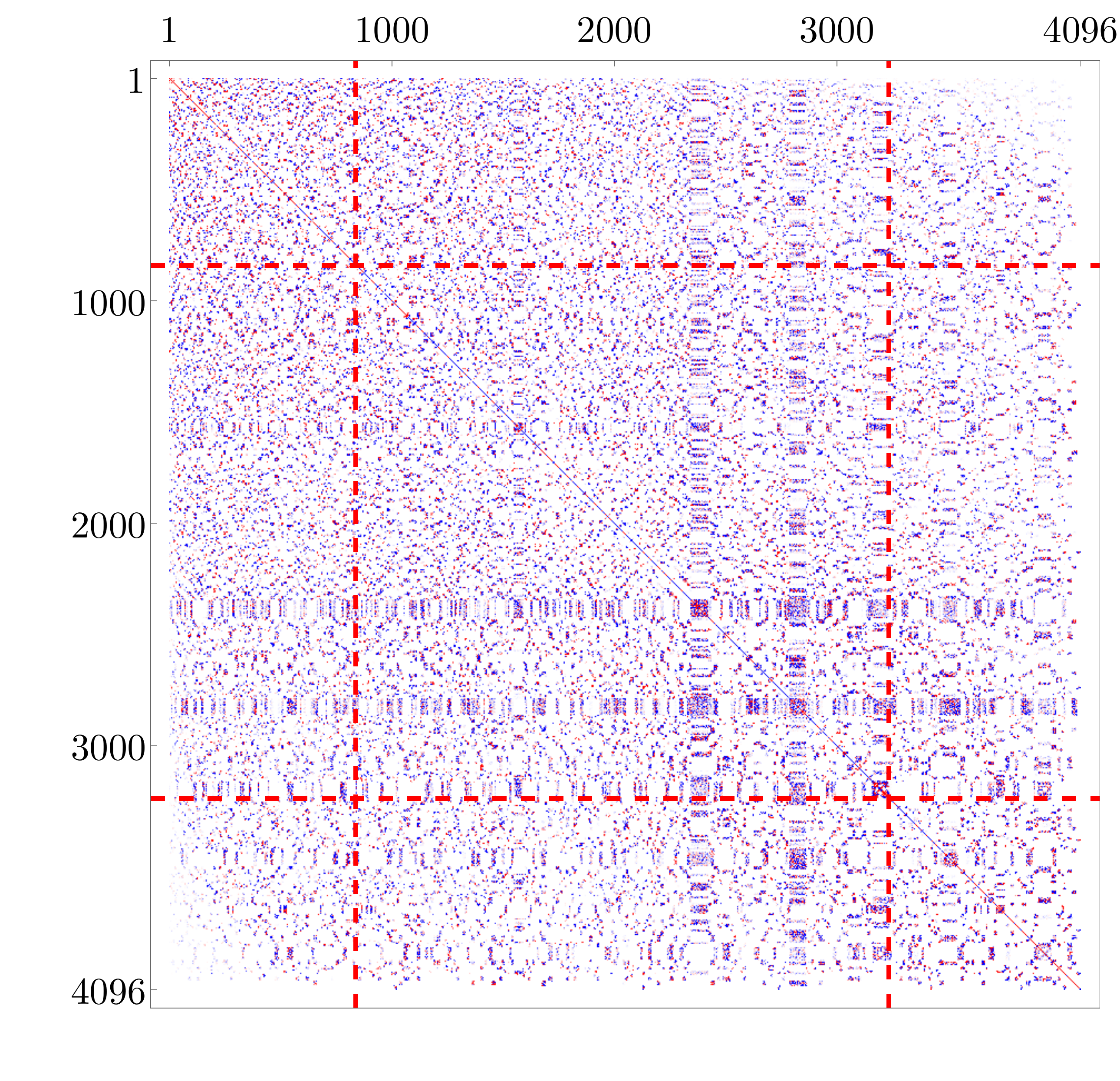}
	\caption{Matrix plot of the pertubation $V$ in the eigenbasis of the unperturbed Hamiltonian $H_0$. Red lines mark the interval of approximately constant DOS, cf. Fig. \ref{dos}.}
	\label{matrixnormal}
\end{figure}

\noindent
The assessment of this finding is twofold: On the one hand, ``natural" initial states
like $\rho_1(0)$ and $\rho_2(0)$ do not necessarily live entirely in an energy window of strictly constant DOS. On the other hand, Fig. \ref{dos} indicates that
the states $\rho_1(0)$ and $\rho_2(0)$ are not completely off such a description. One may thus be inclined to expect at least qualitatively reasonable results from an 
application of the theory presented in Ref. \cite{reimann1}.  Concerning  perturbations $V$, the approach in Ref. \cite{reimann1} strictly speaking makes no restrictions,
except for ``smallness". But the result from Ref. [8] is of statistical nature: To the overwhelming majority of the  ensemble of matrices $V$ that is generatated by drawing 
matrix elements in the eigenbasis of $H_0$ independently at random (according to some probability distribution, which may give rise to some sparseness)
the prediction of Ref. \cite{reimann1} 
(exponential damping) applies. While any $V$ may be viewed as an instance of this set, not all $V$ are equally likely. Again, judging the ``typicality'' 
of some concrete $V$ is hard. However, for a qualitative evaluation of the 
typicality  of the  pertubation $V$ at hand with respect to the above ensemble,
a color-scaled plot of $V$ in the  energy eigenbasis of $H_0$ for $N=12$ is depicted in Fig. \ref{matrixnormal}. The red lines correspond to the energy regime marked 
in Fig. \ref{dos}. Obviously, there is some sparseness, about $2\%$ of all elements differ from zero. Other than that the assessment of this finding is also twofold:
On the one hand, some structure is visible in Fig.  \ref{matrixnormal}. On the other hand, this structure is not sufficient to clearly  identify $V$ as 
particularly untypical.
Hence, again, one may be inclined to expect at least qualitatively reasonable results from an 
application of the theory presented in Ref. [8]. \newline

\noindent

\noindent
The theory advocated in Ref. \cite{richter1} relies on projection operator techniques \cite{breuer} and thus has, in principle, no  formal applicability limit. 
However, as projection operator techniques result in perturbative expansions, concrete predictions going beyond leading order are very hard to obtain \cite{xxx}. 
Even the accurate computation of the leading order requires the knowledge of the detailed form of the matrix depicted in \mbox{Fig. \ref{matrixnormal}} 
has to be taken into account. The simple guess of an exponential damping at sufficiently long times only results under preconditions that are rather similar 
to the ones on which the approach from Ref. \cite{reimann1} is based. The conditions under which the dynamics is well captured by a 
leading order description are technically hard to define and even harder to check. However, there are indications that the sparseness of the matrix depicted 
in Fig. \ref{matrixnormal} threatens the correctness of a leading order calculation \cite{bartsch}.\newline

\noindent
The approach advocated in Ref. \cite{stability} is heuristic and primarily based on some numerical evidence, thus no formal preconditions 
may be formulated so far, cf. Sect. \ref{memkern}.
 However, the numerical examples in  Ref. \cite{stability} to which this scheme applies do feature unperturbed Hamiltonians 
with constant DOS, weak perturbations (small  $\lambda$) 
and initial states of the type  $\rho_1(0)$. Furthermore the $V$'s in the examples in Ref. \cite{stability} are matrices whose elements, in the
eigenbasis of the observable, are independently drawn at random according to some probability distribution.
As already mentioned in Sect. \ref{memkern}, in contrast to Refs. \cite{reimann1,richter1}, the approach in  Ref. \cite{stability} 
takes the structure  of $V$ in the eigenbasis of the observable (here $S_q^z$) rather than of $H_0$ into account. 
Only if the latter approximately commute, i.e. $[V,A]\approx 0$, the prediction computed as described in Sect. \ref{memkern} applies. 
For our setup we indeed have $[V,A]= 0$, cf. Sect. \ref{setup}.\newline

\noindent
We now embark on the announced bold comparison of the perturbed dynamics with the predictions from the three theories.\newline
\noindent
Firstly, note that for both initial states the perturbed curves lie above the unperturbed one, i.e. the stronger the perturbation the slower the relaxation occurs.
Thus, theories predicting a damping of the unperturbed dynamics are not a viable option in this case. Without any further quantitative analysis this already renders the 
predictions from  Ref. \cite{reimann1} and Ref. \cite{richter1} qualitatively unsuitable. Moreover, it can be shown (at least for the first initial state) 
that all curves must feature zero
slope at $t=0$. An exponential damping 
(with a constant damping factor) would always change the slope at $t=0$ to a non-zero value. A time-dependent damping factor $\Gamma(t)$ with $\Gamma(0)=0$
(as employed in \mbox{Ref. \cite{richter1}}) at least preserves the zero slope at $t=0$. \newline
\noindent
These findings suggest that the pertubation $V$ is indeed one of the mathematically extremely untypical members of the ensemble considered in Ref. \cite{reimann1},
even though the matrix visualization in Fig. \ref{matrixnormal} does not necessarily indicate this. However, 
even though $V$ is untypical with respect to an ensemble of random matrices, it is a physically simple, common  pertubation consisting of 
standard spin-spin interactions.
\newline




\begin{figure}[t!]
\centering
  	\includegraphics[width=0.45\textwidth]{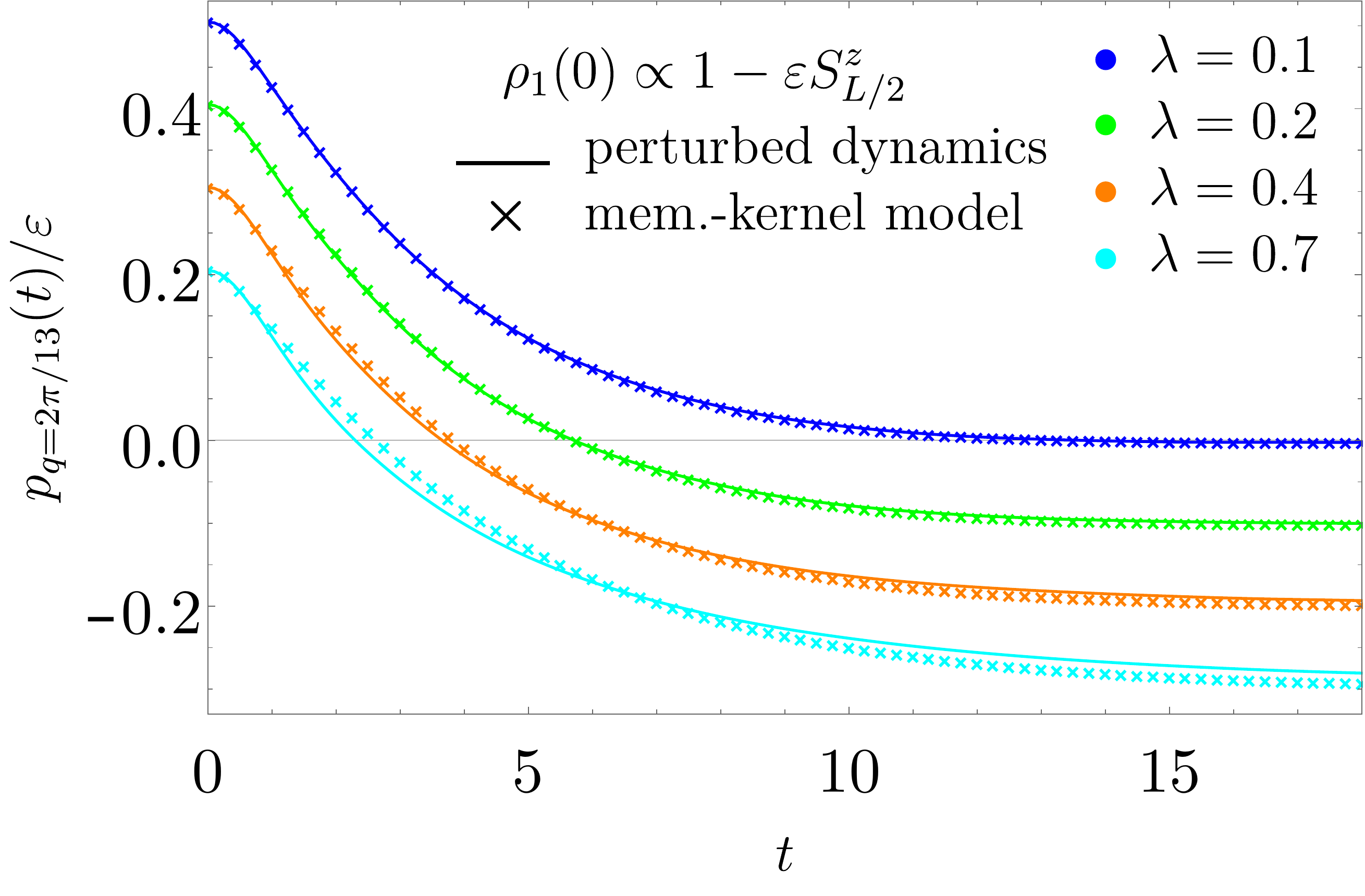}
	\caption{Slowest mode with \mbox{$q=2 \pi/13$} of the infinite-temperature initial state $\rho_1(0)$ depicted for various pertubation strengths. Solid lines represent the original data from Fig. \ref{dyn}, vertically shifted in steps of $-0.1$ for better visibility. Crosses indicate the data obtained from an exponentially damped memory-kernel.}
	\label{dynmem}
\end{figure}
\noindent
The failure of the scheme presented in  Ref. \cite{richter1} indicates that the $V$ at hand  does not allow for a leading order 
truncation of the projective scheme 
employed therein, not even for very small $\lambda$. This leaves the memory-kernel model from Ref. \cite{stability} as a the only 
feasible theory to describe the observed behavior.  \newline

\noindent
 In the following, to test the approach from Ref. \cite{stability}, we apply the memory-kernel ansatz to the two unperturbed dynamics (red curves in Fig. \ref{dyn} and Fig. \ref{dyn2}). The damping constant $\gamma$ from Eq. \eqref{damp} functions as a fit parameter and is optimized such that the $L^2$-error of the two curves in question (perturbed dynamics and memory-kernel prediction) is minimized.
\begin{figure}[b!]
\centering
  	\includegraphics[width=0.45\textwidth]{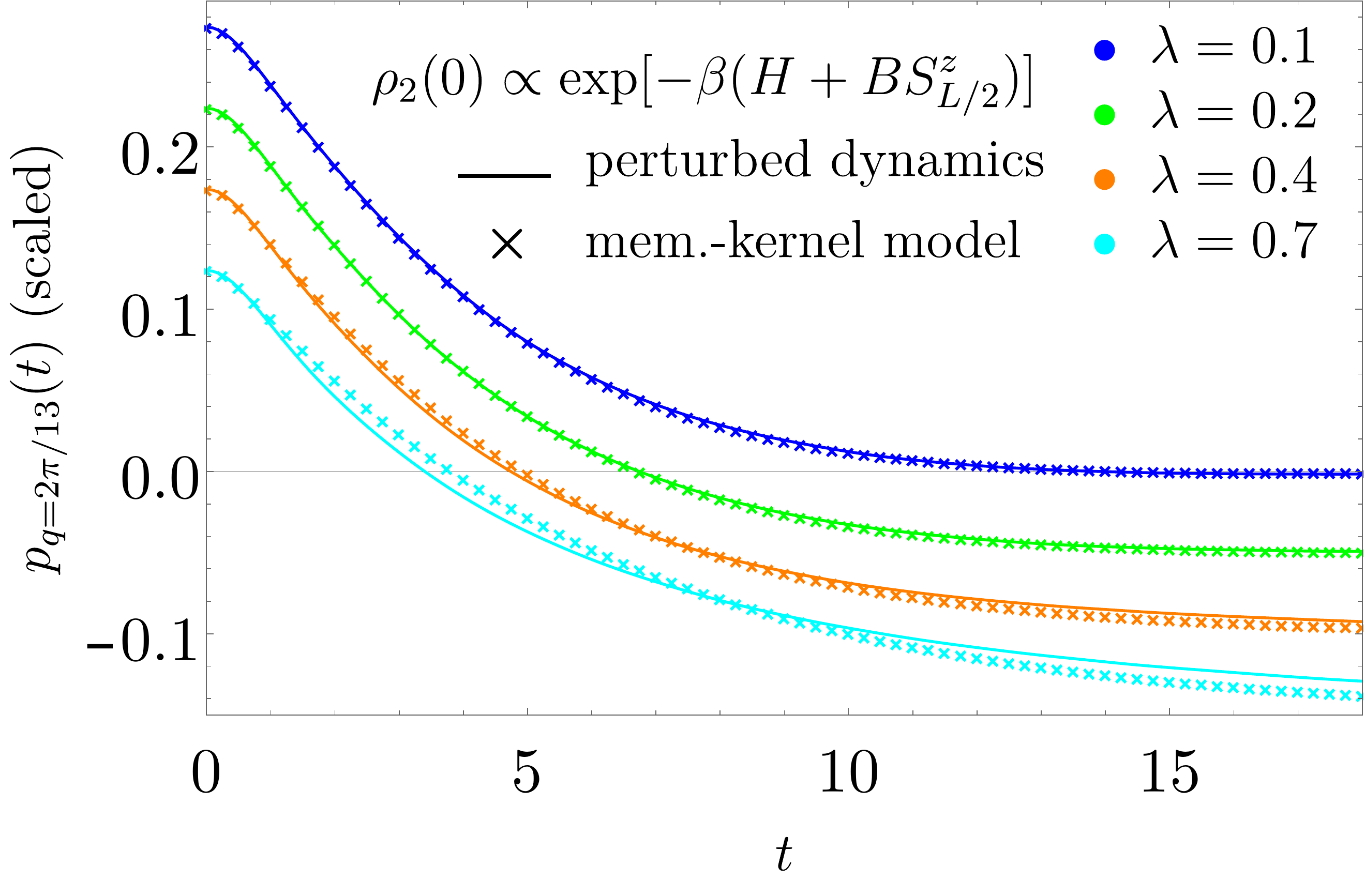}
	\caption{Slowest mode with $q=2 \pi/13$ of the Gibbsian initial state $\rho_2(0)$ with $\beta=0.1$ and $B=5.0$ depicted for various pertubation strengths. Solid lines represent the original data from Fig. \ref{dyn2}, vertically shifted in steps of $-0.05$ for better visibility. Crosses indicate the data obtained from an exponentially damped memory-kernel.}
	\label{dynmem2}
\end{figure}
\newpage

\noindent
  For the infinite-temperature initial state $\rho_1(0)$ the results are depicted in \mbox{Fig. \ref{dynmem}}, for the Gibbsian initial state $\rho_2(0)$ the results are depicted in Fig. \ref{dynmem2}.
Each curve is vertically shifted to avoid clutter.
For the infinite-temperature initial state $\rho_1(0)$ and weak pertubations ($\lambda =0.1$ and $\lambda= 0.2$) the memory-kernel model seems to perfectly capture the modified dynamics. For stronger pertubations ($\lambda = 0.4$ and $\lambda = 0.7$) there are deviations visible, e.g. for short times ($t \sim 2$) the memory-kernel prediction for $\lambda=0.7$ overshoots the perturbed dynamic while for longer times $t \gtrsim 10$ it undershoots. For the Gibbsian initial state $\rho_2(0)$ the qualitative behavior remains the same as for the first initial state. For weak pertubations the memory-kernel ansatz captures the modifications due to the pertubation extremly well. For stronger pertubations there are again more noticeable deviations. However, it comes as no surprise that the memory-kernel model looses potency in the strong pertubation regime, since it was originally conceived to describe the alteration of dynamics due to weak pertubations.
\vspace*{-9px}
\section{Summary and conclusion}
\label{sumcon}
\vspace*{-9px}
In the paper at hand we numerically analyzed the applicability of three theories
predicting the generic impact of Hamiltonian perturbations on expectation value dynamics to a Heisenberg spin ladder. To this end, we numerically calculated the time-dependent spatial distribution of the magnetization along the ladder for 
various pertubation strengths. We focused on a particular perturbation that commutes with the observable, e.g. the considered perturbation $V$ 
consisting of  $S^zS^z$-couplings on the ladder diagonals commutes with the observed spatial magnetization distribution. We consider both, infinite and finite
temperatures. Two out of three scrutinized theories feature in principle well defined conditions for their applicability \cite{reimann1,richter1}, a third one is 
rather heuristic \cite{stability}. One of the theories  
with well defined conditions \cite{reimann1} only predicts the overwhelmingly likely behavior with respect to a hypothetical, large, ``random matrix ensemble''    of in principle possible perturbations $V$. Only the 
heuristic theory takes the the commutativity of the  observable and the perturbation as a specifically relevant structural feature into account. 
It turns out to be hard to judge {\em a priori} whether or not the  concrete spin ladder example falls into the realm of applicability of the two theories with well 
defiend condtions. However, direct comparison of the theoretical predictions with the numerically computed results clearly shows that both theories fail even qualitativley.
This suggests that, while the the considered $V$ is very common from a physical point of view, it must be very rare and exotic with respect
to the above random matrix ensemble. Only the heuristic theory was found to yield good  results for weak perturbations (and acceptable results for strong perturbations).
This indicates that the commutator of $V$ with the observable is a specifically relevant structural feature that should be taken into account. 
A survey of the three theories for perturbations $V$ that do not commute with the observable is left for further research.
\vspace*{-9px}
\begin{acknowledgments} 
\vspace*{-9px}
We thank P. Reimann and L. Dabelow for fruitful discusssions on this subject.
This work was supported by the Deutsche Forschungsgemeinschaft (DFG)
within the Research Unit FOR 2692 under Grant No. 397107022.
\end{acknowledgments}
\bibliography{literature}

\end{document}